# Strong Gravitational Lensing in a Brane-Wolrd Black Hole

Guoping Li[1*]    Biao Cao[2**]    Zhongwen Feng[1]    Xiaotao Zu[1]

[1]School of Physical Electronics, University of Electronic Science and Technology of china,

Chengdu, Sichuan 610054, China

[2]Institute of theoretical physics, China West Normal University, Nanchong, Sichuan 637009, china

**Abstract:** Adopting the strong field limit approach, we investigated the strong gravitational lensing in a Brane-Word black hole, which means that the strong field limit coefficients and the deflection angle in this gravitational field are obtained. With this result, it can be said with certainly that the strong gravitational lensing is related to the metric of gravitational fields closely, the cosmology parameter $\alpha$ and the dark matter parameter $\beta$ come from the Brane-Wolrd black hole exerts a great influence on it. Comparing with the Schwarzschild-AdS spacetime and the Schwarzschild-XCMD spacetime, the parameters $\alpha, \beta$ of black holes have the similar effects on the gravitational lensing. In some way, we infer that the real gravitational fields in our universe can be described by this metric, so the results of the strong gravitational lensing in this spacetime will be more reasonable for us to observe. Finally, it has to be noticed that the influence which the parameters $\alpha, \beta$ exerted on the main observable quantities of this gravitational field is discussed.

## 1 Introduction

In the view of the gravitational theory, photons would be deviated from their straight path when they pass close to a compact object. The phenomena of the light deflection called the gravitational lensing, the compact object casing the detectable deflection named the gravitational lens. In the early 20[th] century, the astronomical phenomenon of the gravitational lensing has been tested[1]. Now, the gravitational lensing has most important status and special significance in the cosmology and the astronomy. First, the properties of the gravitational lens are related to the whole universe, so we can constraint the Hubble constant and the cosmological constant with the analysis of the gravitational lensing. Secondly, gravitational lensing plays important roles of the astronomical telescope, and it can't only increase the brightness of the image which the source produced, but also can help us extract the information about stars which are too dim to be observed. Finaly, the gravitational lensing also has some relations to the dark matter, the quasiconformal modes and the extra dimensions. At the beginning of the 21[th] century, the strong gravitational lensing, the weak gravitational lensing and the micro gravitational lensing all have been researched carefully[2-7]. It is generally recognized that the strong gravitational lensing can be used to study the background objects which is far away form us[2], the weak gravitational lensing has the important significance to the cosmology[3], and the micro gravitational lensing can use to realize the mass distributions of the fixed stars[4-7]. As a result, the study of the gravitational lensing has become one of the most popular research fields in the cosmology, astronomy and the theoretical physics. In 2002, V. Bozza obtained a successful achievement in the research of the strong gravitational lensing[8-9]. He put forward a reliable method to calculate the light deflection angle in a spherically symmetric spacetime with the Virbhadra-Ellis lens equation. The result indicates that the deflection angle diverges logarithmically when the light ray is close to the strong gravitational fields. Latter, Chen and Wei *et al.* studied the strong gravitational lensing of the squashed Kaluza-Klein black hole and the Kerr black hole pierced by a cosmic string with this method and the similar results has been obtained[8-9]. On the other hand, lots of work about the strong gravitational lensing has been studied[37-43], in that this method is a reliable method to study the strong gravitational lensing[10-23]. However, the research of the gravitational lensing remained need us to continue perfecting it. At present, we expect that the gravitational lensing can play a more important role in the theoretical physics, the cosmology and the astronomy and so on.

---
[*] GuoPing Li:  li_gp2009@163.com  
[**] Corresponding author(Biao Cao): caobiaocwnu@163.com

In 1933, Zwicky put forward the concept of the dark matter with their astro-observations[24]. Then, in view of the present research, it is hardly too much to say that the dark matter exists and it is a truth for us. More and more evidences indicate that the dark matter take the occupancy about 23% in our universe. However, the occupancy of the ordinary matter is only 4% which is much less than the dark matter. Despite that the study of the dark matter has obtained some progresses, there still have many problems and further more researches need us to study. Obviously, the dark matter research has become one of the most important research fields in the theory physics[25-28]. In the early 1990s, the famous Kaluza-Klein theory has been put forward by M.J.Duf, and the extra dimension faded into people's sight along with it. However, the extra dimension has no detectability, in that the idea of the extra dimension faded away. Then, the recent research shows that the string theory may be a candidate for the unified theory, and the Brane-World theory which promoted the extra dimension theory greatly has developed with the string theory[30-31,36]. The Brane-World theory is that people live in a piece of four-dimensional brane that set into the higher-dimensional brane; on the four-dimensional brane, three physical interactions are restricted with the exception of the gravitational interaction, which means that the gravity can propagate free in the higher-dimensional brane only. Also, the Brane-World has the Thick-Brane type and Thin-Brane type, and the theoretical models of the Brane-World are various[32-35]. In brief, it can not only enriched the extra dimensional theory greatly but also explained some physical problems very well by this idea. At present, the Brane-World theory has attracted more and more attention of physical scientists.

In this paper, we studied the strong gravitational lensing in a Brane-Word black hole with the strong field limit approach, and the strong field limit coefficients $\left(\bar{a},\bar{b}\right)$ and the deflection angle $\alpha(\theta)$ are obtained. Discussing with the parameter $\alpha$ $\beta$, we analyzed the influence that parameters $\alpha, \beta$ have on the gravitational lensing. Various kinds of evidences indicate that the real black hole in our universe should include the dark matter, so the black hole in the centre of our galaxy or the others can be described by this black hole. Hence, the metric of the Brane-Word black hole in our real universe is more realistic than other metrics and the gravitational lensing results come from it will be more significance. Finally, comparing with the Schwarzschild-AdS spacetime and the Schwarzschild-XCMD spacetime, I will conclude by saying that the parameters $\alpha, \beta$ have same effects on the observable quantities in carved spacetime.

## 2 the strong gravitational lensing of the Brane-World black hole

In 2008, According to the total action of the carved spacetime[29],

$$S = \frac{1}{2\alpha_*} \int_{\mathcal{M}} d^m X \sqrt{-\mathcal{G}} \left( \mathcal{R} - 2\Lambda^{(b)} \right) + \int_{\Sigma} d^4 x \sqrt{-g} \left( \mathcal{L}_{surface} + \mathcal{L}_m \right). \tag{1}$$

Using the complicated computation and simplification, the Einstein field equation in the Brane-Wolrd has been obtained by M.Heydari-Fard. So the vacuum field equation can be written as follow,

$$G_{\mu\nu} = -\Lambda g_{\mu\nu} + Q_{\mu\nu} - \varepsilon_{\mu\nu}, Q_{\mu\nu} = \left( KK_{\mu\nu} - K_{\mu\alpha} K_{\nu}^{\alpha} \right) + \frac{1}{2} \left( K_{\alpha\beta} K^{\alpha\beta} - K^2 \right) g_{\mu\nu}. \tag{2}$$

Where, $Q_{\mu\nu}$ is the independent conservation quantity in the Brane-Wolrd, and it represents the energy-momentum tensor which is related to the dark matter[29]. With the vacuum field equation, a spherically symmetric black hole solution with dark matter in the Brane-Wolrd is obtained,

$$ds^2 = -f(r) dt^2 + f(r)^{-1} dr^2 + r^2 \left( d\theta^2 + \sin^2 \theta d\phi^2 \right). \tag{3}$$

$$f(r) = 1 - \frac{1}{r} - \alpha^2 r^2 - 2\alpha\beta r - \beta^2. \tag{4}$$

Here, $\alpha$ is a metric parameter which is related to the cosmological constant. If choose the condition $\beta = 0$, so $f(r) = 1 - \frac{1}{r} - \alpha^2 r^2$ and the spacetime can be degraded into the Schwarzschild-AdS(4) solution; If the situation $\alpha = 0$ is satisfied, $f(r) = 1 - \frac{1}{r} - \beta^2$ and the Schwarzschild-XCDM solution is obtained; and if we consider the condition $\alpha = \beta = 0$, the spacetime can be degraded into Schwarzschild black hole finally. In this paper, we will calculate the strong gravitational lensing in the Brane-Wolrd black hole with V.Bozza's viewpoint. In order to calculate conveniently, we define the Schwarzschild radius $r_s = 2M = 1$ as the measure of distances and consider the specific situation that the orbital plane of the photon is $\theta = \pi/2$ only. Then, in the standard coordinates, the functions in the metric become,

$$ds^2 = A(r) dt^2 - B(r) dr^2 - C(r) d\varphi^2. \tag{5}$$

$$A(r) = f(r), B(r) = f(r)^{-1}, C(r) = r^2. \tag{6}$$

In the strong gravitational lensing, it's geometric principle is that when a beam of light from the source(S) pass close to the gravitational lens(L), the light ray would be deviated. For studying the strong gravitational lensing more convenient, the situation that the source (S), the gravitational lens (L) and the observer are in the same orbital plane has been considered. According to the geometric principle, the relationship between the image and the source can be described by the

Virbhadra-Ellis lens equation which is $tan\tilde{\theta} - tan\tilde{\beta} = D_{LS} / \left[ tan\tilde{\theta} + tan(\tilde{\alpha} - \tilde{\theta}) \right] \times D_{OS}$ [44]. The parameter $\tilde{\beta}$ represents the angle between the source and the lens, and $\tilde{\theta}$ is the angle between the image and the lens, $\tilde{\alpha}$ is the deflection angle. The distance from the observer to the lens is $D_{OL}$, and $D_{LS}$ is the distance between the source and the lens, $D_{OS}$ represents the distance from the source to the observer. Meanwhile, the condition of the strong gravitational lensing can be satisfied, which is $D_{OL} + D_{LS} = D_{OS}$. From the geodesics equation, the angular deflection of the photon in the $\varphi$ direction is $d\varphi = dr \sqrt{B} / \left( \sqrt{C} \sqrt{(C \times A_0)/(C_0 \times A) - 1} \right)$ [46-47], so the deflection angular can be written as[46],

$$\alpha(r_0) = I(r_0) - \pi, \quad I(r_0) = \int_{r_0}^{\infty} \frac{2\sqrt{B}}{\sqrt{C}\sqrt{\frac{C}{C_0}\frac{A_0}{A} - 1}} dr. \quad (7)$$

Obviously, the deflection angle depends on the closest approach distance directly. With the $r_0$ decreasing, whereas the deflection angle increasing, when the $r_0$ decrease to a certain value, the deflection angle will exceed $2\pi$. As a result, the photon will arrive at observer after it took a complete loop around the strong gravitational fields. On the other hand, the deflection angle will diverge if the condition $r_0 = r_m$ is satisfied. Then, the photon will be captured by the strong gravitational fields completely. Here, the $r_m$ is the radius of photon sphere which is from the photon sphere equation $C'(r)/C(r) = A'(r)/A(r)$ [45,48]. Substituting Eq.(4.6) into the photon sphere equation, so,

$$r_m = \frac{1 - \beta^2 - \sqrt{1 - 6\alpha\beta - 2\beta^2 + \beta^4}}{2\alpha\beta}, r_{m1} = \frac{1 - \beta^2 + \sqrt{1 - 6\alpha\beta - 2\beta^2 + \beta^4}}{2\alpha\beta}. \quad (8)$$

From Eq.(8), the photon sphere equation has two roots $(r_m, r_{m1})$. In order to match the situation of Schwarzschild spacetime, we expand the $(r_m, r_{m1})$ at the position of $\alpha=0, \beta=0$ and find that $r_m$ is the radius of photon sphere in the Brane-Word black hole. For different gravitational fields, the radiuses are different, so the Schwarzschild black hole and the Schwarzschild-AdS black hole have the same value $r_m^S = r_m^{S-AdS} = 1.5$; For the Schwarzschild-XCDM black hole, the radius of photon sphere is $r_m^{S-X} = 3/(2 - 2\beta^2)$; But for the stationary axisymmetric black hole, it only has a similar concept about the radius of photon sphere, and the deflection angle will also be infinity when the condition $r_0 \to r_m$ is satisfied. At the point of V.Bozza, we defined the variate as follow,

$$z = 1 - r_0/r. \tag{9}$$

So,

$$I(r_0) = \int_0^1 R(z, r_0) f(z, r_0) dz. \tag{10}$$

Where,

$$R(z, r_0) = \frac{2r^2 \sqrt{A(r)B(r)C(r_0)}}{r_0 C(r)} = 2. \tag{11}$$

$$f(z, r_0) = \frac{1}{\sqrt{A(r_0) - A(r_0)\frac{C(r_0)}{C(r)}}}. \tag{12}$$

The subscript 0 shows that the function is evaluated at $r_0$, and $R(z, r_0)$ is regular for all values of $z$ and $r_0$, while $f(z, r_0)$ divers for $z \to 0$. To solve the integral Eq.(10), we split it into two pieces,

$$I_D(r_0) = \int_0^1 R(0, r_m) f_0(z, r_0) dz. \tag{13}$$

$$I_R(r_0) = \int_0^1 (R(z, r_0) f(z, r_0) - R(0, r_m) f_0(z, r_0)) dz. \tag{14}$$

In Eq.(12), the function $f(z, r_0)$ is divergence. In order to find out the order of divergence of the integrand, we expand the argument of the square root in $f(z, r_0)$ to the second order in $z$,

$$f(z, r_0) = \frac{1}{\sqrt{M(r_0)z + N(r_0)z^2}}. \tag{15}$$

Here,

$$M(r_0) = 2 - 2\beta^2 - \frac{3}{r_0} - 2\alpha\beta r_0. \tag{16}$$

$$N(r_0) = \beta^2 - 1 + \frac{3}{r_0}. \tag{17}$$

When $r_0$ approach to $r_m$, so we have $M(r_0) = 0$ and the divergence is $z^{-1}$ which makes the integral diverge. The expression of the deflection angular is,

$$\alpha(\theta) = -\bar{a} Log\left(\frac{\theta D_{OL}}{u_m} - 1\right) + \bar{b} + \mathcal{O}(u - u_m). \tag{18}$$

And,

$$\bar{a} = \frac{1}{\sqrt{N(r_m)}}. \tag{19}$$

$$\bar{b} = -\pi + b_R + \bar{a} Log\left(\frac{r_m^2 \left[C''(r_m) A(r_m) - A''(r_m) C(r_m)\right]}{u_m \sqrt{A^3(r_m) C(r_m)}}\right). \tag{20}$$

$$b_R = I_R(r_m) \quad , \quad u_m = \frac{r_m}{\sqrt{A(r_m)}}. \tag{21}$$

The parameters $(\bar{a}, \bar{b})$ are the strong field limit coefficients which are all related to the radius of photon sphere $r_m$, and $u_m$ represent the impact parameter with the angular momentum. Here, we first discuss the special situation of the Brane-Wolrd black hole which is the Schwarzschild-AdS(4) spacetime and the Schwarzschild–XCDM spacetime, this is very important for the study of the gravitational lensing in the Brane-Wolrd black hole. For the Schwarzschild-AdS(4) spacetime and the Schwarzschild spacetime, the integral parts are all $b_R^{S-AdS} = b_R^S = 2Log\left(6(2-\sqrt{3})\right)$, and we can obtain the angular deflection $\alpha(\theta)$ of the Schwarzschild-AdS(4) spacetime and Schwarzschild–XCDM spacetime,

$$\alpha(\theta)^{AdS} = -Log\left(\frac{0.0005}{3}\sqrt{\frac{4-27\alpha^2}{12}}\right) - \pi + 2Log\left(6(2-\sqrt{3})\right)$$
$$+Log\frac{28}{4-27\alpha^2} \tag{22}$$

$$\alpha(\theta)^{XCDM} = -\frac{1}{\sqrt{1-\beta^2}} Log\left(\frac{0.0005 * 2\sqrt{3}\sqrt{(1-\beta^2)^3}}{9}\right) + Log\left(\frac{3}{1-\sqrt{6}}\right)^2$$
$$+\frac{1}{\sqrt{1-\beta^2}} Log 6 - \pi \tag{23}$$

Using the Eq.(22) and Eq.(23), we plot the graph of the two special situations of the Brane-Wolrd black hole with the condition $u = u_m + 0.0005$, the following graphs which are obtained as,

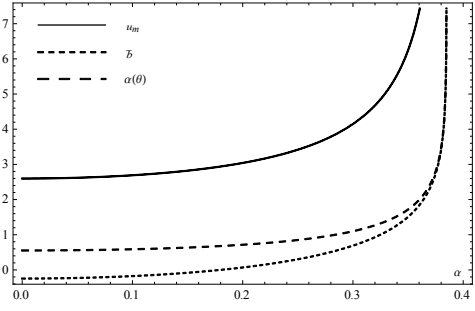 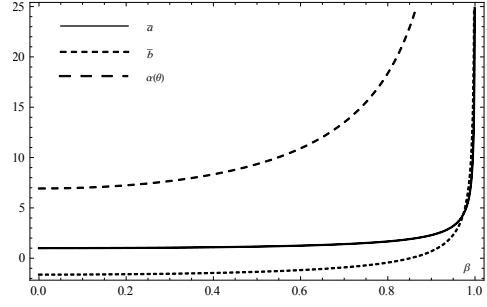

FIG.1: Schwarzschild-AdS(4) spacetime       FIG.2: Schwarzschild-XCDM spacetime

The FIG.1 and the FIG.2 show that the metric parameters $(\alpha,\beta)$ have a significant influence on the strong gravitational fields. For the Schwarzschild-AdS spacetime, the cosmology parameter $\alpha$ paly a important role in the strong gravitational lensing. With the increasing of the parameter $\alpha$, the impact parameter $u_m$, the gravitational lenging coefficient $\bar{b}$ and the deflection angle $\alpha(\theta)$ also increased gradually; however, another gravitational lenging coefficient $\bar{a}$ and the radius of photon sphere $r_m$ have the same value with the Schwarzschild spacetime. For the Schwarzschild-XCDM spacetime, the influence that the parameter $\beta$ has on the gravitational lensing can be described as: the impact parameter $u_m$, the gravitational lenging coefficients $(\bar{a},\bar{b})$ and the deflection angle $\alpha(\theta)$ should increase with the increasing of parameter $\beta$. In a word, the parameters $(\alpha,\beta)$ affect the gravitational lenging of the two special spacetime in the Brane-Wolrd black hole seriously.

For the strong gravitational fields $f(r)=1-\frac{1}{r}-\alpha^2 r^2 - 2\alpha\beta r - \beta^2$, the following results are obtained with the detailed calculations,

$$b_R = 2 \times Log\left(\frac{4\Delta}{2\Delta - \frac{1}{r_m} + 2\sqrt{\Delta^2 - \frac{\Delta}{r_m}}}\right), \bar{a}=\frac{1}{\sqrt{\Delta}}, \Delta=\beta^2-1+\frac{3}{r_m}. \qquad (24)$$

$$u_m = \frac{r_m}{\sqrt{1-\frac{1}{r_m}-\alpha^2 r_m^2 - 2\alpha\beta r_m - \beta^2}}, r_m = \frac{1-\beta^2+\sqrt{1-6\alpha\beta-2\beta^2+\beta^4}}{2\alpha\beta}. \qquad (25)$$

$$\bar{b} = -\pi + \frac{1}{\sqrt{\Delta}} Log\left(2 + \frac{2\alpha^2 r_m^3 + 2}{r_m - 1 - \alpha^2 r_m^3 - 2\alpha\beta r_m^2 - r_m\beta^2}\right)$$

$$+ 2Log\left(\frac{2\sqrt{\Delta}}{\sqrt{\Delta - \frac{1}{r_m}} + \sqrt{\Delta}}\right)^2. \qquad (26)$$

$$\alpha(\theta) = -\frac{1}{\sqrt{\Delta}} Log\left(\frac{u}{u_m} - 1\right) + \bar{b} + \mathcal{O}(u - u_m). \qquad (27)$$

In order to understand the influence intuitively that the parameters $(\alpha, \beta)$ have on the gravitational lensing, so we plotted the graph of the gravitational lensing as Figs.3-5,

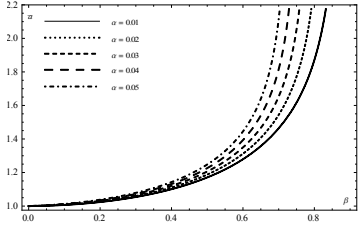 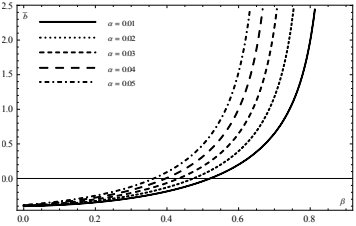 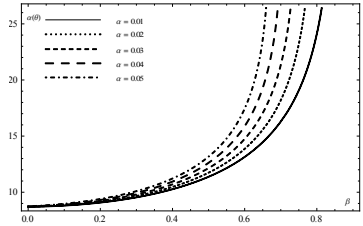

FIG.3: $\bar{a}$          FIG.4: $\bar{b}$          FIG.5: $\alpha(\theta)$

From the FIGS.(3-5), we find that: with the increasing of the cosmology parameter and the dark matter parameter, not only the gravitational lenging coefficients, but also the angular deflection increased gradually. So, the influence that the parameters $(\alpha, \beta)$ have on the strong gravitational lensing is very important.

### 3 The observable quantities of the strong gravitational lensing

The Eq. (24-27) and FIG. (3-5) is mathematical expression of the strong gravitational lensing in Brane-Word black hole. With the map of geometric structure, the lens equation can be simplified as Eq.(37) when the condition $(\tilde{\beta}, \tilde{\theta} \sim 0)$ is satisfied.

$$\tilde{\beta} = \tilde{\theta} - \frac{D_{LS}}{D_{OS}} \Delta\alpha_n. \qquad (28)$$

Where, $\Delta\alpha_n = \tilde{\alpha} - 2n\pi$, $n$ is the number of loops that the photon surround the gravitational fields. Then, we can obtain the position of the image $\tilde{\theta}_n$ and the magnification $\mu_n$ of the n$^{\text{th}}$ image,

$$\tilde{\theta}_n = \tilde{\theta}_n^0 + \frac{u_m e_n \left(\tilde{\beta} - \tilde{\theta}_n^0\right) D_{OS}}{\bar{a} \tilde{\beta} D_{LS} D_{OL}}. \qquad (29)$$

$$\mu_n = e_n \frac{u_m^2 (1+e_n) D_{OS}}{\bar{a} \tilde{\beta} D_{OL}^2 D_{LS}}. \tag{30}$$

Here,

$$\widetilde{\theta_n^0} = \frac{u_m (1+e_n)}{D_{OL}}, e_n = exp\left(\frac{\bar{b} - 2n\pi}{\bar{a}}\right). \tag{31}$$

From the Eq.(31), the approximate expression $\mu_n \sim e^{-n}/(\tilde{\beta} D_{OL}^2)$ indicates that, the brightness of the image decreased with the exponent, and the first level of the image is the brightest level; for the same level, with the distance $D_{OL}$ increasing, the brightness of the image decreased; on the other hand, if the smaller the angle $\tilde{\beta}$, the more bright of the image. When the levels $n$ approach to $\infty$, the relationship between the impact parameter and the asymptotic position of the relativistic image is,

$$\widetilde{\theta_\infty} = \frac{u_m}{D_{OL}}. \tag{32}$$

For the convenience, we usually separate the outermost image from the other images, which is $\widetilde{\theta_1}$, and the surplus images are regarded as a group $\widetilde{\theta_\infty}$, so the angular separation is,

$$s = \widetilde{\theta_1} - \widetilde{\theta_\infty} = \theta_\infty \cdot exp\left(\frac{\bar{b} - 2\pi}{\bar{a}}\right). \tag{33}$$

Then, the ratio of luminous flux between the first image and other ones can be expressed as,

$$\tilde{r} = \mu_1 \bigg/ \sum_2^\infty \mu_n = exp(2\pi/\bar{a}). \tag{34}$$

Where, the position of image $(\widetilde{\theta_\infty})$, the angular separation $(s)$ and the ratio of luminous flux $(\tilde{r})$ are all observable quantities. With the measure of those observable quantities, the limit coefficient of the strong gravitational lensing and the impact parameter are obtained. Therefore, we can test the validity of the strong gravitational lensing model through the analysis between the observable quantities and the theoretical value.

Substituting Eq.(24,26) into Eq.(32-34), the variation of the three observable quantities with the parameter $(\alpha, \beta)$ is,

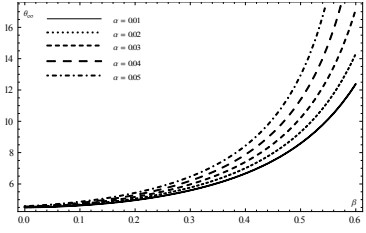 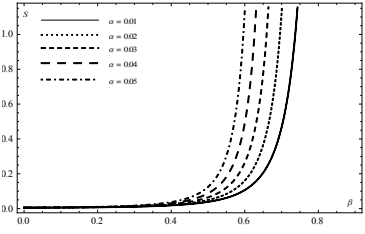 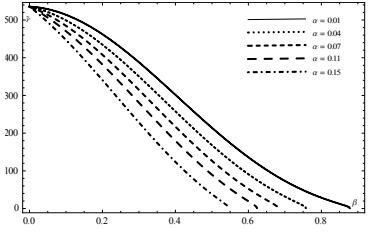

FIG.6: $\widetilde{\theta_\infty}$      FIG.7: $S$      FIG.8: $\tilde{r}$

Obviously, if the value of the parameters $(\alpha,\beta)$ increased, the position of image and the angular separation will increase; On the contrary, the luminous flux will reduce with the decreased of parameter $(\alpha,\beta)$. From here we see that the parameters $(\alpha,\beta)$ are very important to the strong gravitational lensing in our universe and it is worthy of being study for us.

## 4 Conclusions

In this paper, we studied the strong gravitational lensing in a Brane-Word black hole with the V.Bozza's method and obtained the strong field limit coefficients and the deflection angle in this gravitational field. The result shows that the metric parameters have influence on the strong gravitational lensing greatly. Because of the present study about the dark matter and the Brane-Word theory, the metric of the Brane-Word black hole is closer to the real black hole in our universe, so the results of the strong gravitational lensing is more reasonable.

Firstly, let's focus on the FIG.(1-8) of the Brane-Word black hole: with the cosmology parameter and the dark matter parameter increasing, the strong fields limit coefficients and the deflection angle increased; also, the position of image and the angular separation will increase if the parameter $(\alpha,\beta)$ increased; however, the luminous flux will reduce with parameter $(\alpha,\beta)$'s increase. From this we inferred those information: the existence of the dark matter and the cosmology constant cause the enhancement of the gravitational fields, so the strong fields limit coefficients, the deflection angle, the position of image and the angular separation increased; on the other hand, the captured ratio of which the black hole capture photons will increase with the rise of the gravitational fields, so the luminance of the first level image decrease more quickly than the surplus images. Secondly, because the extent of the influence which the parameters $(\alpha,\beta)$ have on the strong gravitational lensing is quite deeply, and we infer that the parameters $(\alpha,\beta)$ are the important parts the same as the mass, the charge and the angular momentum of the Brane-Word black hole. Thirdly, the precision of astronomical observation technology is more and more maturely, therefore the results in this paper may be verified totally by the astro-observation. What is more, this result can not only be used to distinguish the spacetimes from the Brane-Word black hole, but also provides a method for estimating that whether the spacetime include the dark matter or not. Finaly, using the FIG.(1-5), we compared the Brane-Word spacetime with the Schwarzschild-AdS spacetime and the Schwarzschild-XCMD spacetime, and got a conclusion that the parameters $\alpha,\beta$ has the same effect on the gravitational lensing parameters.

Recently, the researches on the dark matter, the strong gravitational lensing and the Brane-Word theory have attracted more and more attentions of theory physicists. It has become one of the most important issues in the astronomy as well as cosmology. At present, we only take a preliminary discussion about the dark matter, the Brane-Word theory and the strong gravitational lensing, and we will pay more attention to this issue in the future work.

**Acknowledgement**：This work was supported by the National Foundation of china under Grant NO.11178018.